\documentclass[twocolumn,aps,amsmath,amssymb,showkeys,floatfix,prl,superscriptaddress]{revtex4}

\usepackage{bbm}
\usepackage[dvips]{graphicx}
\usepackage{color}
\usepackage{amsmath,amssymb,latexsym}

\newcommand{\e}{\mathbf{e}}

\newcommand{\n}{\mathbf{n}}

\newcommand{\q}{\mathbf{q}}
\renewcommand{\r}{\mathbf{r}}
\renewcommand{\t}{\mathbf{t}}

\renewcommand{\r}{\mathbf{r}}

\renewcommand{\P}{\mathbf{P}}

\newcommand{\R}{\mathcal{R}}

\newcommand{\chlamy}{\emph{Chlamydomonas}}

\begin{document}

\title{   
In-phase and anti-phase flagellar synchronization by basal coupling
}

\author{Gary S. Klindt}
\affiliation{Max Planck Institute for the Physics of Complex Systems, 01187 Dresden, Germany}
\author{Christian Ruloff}
\affiliation{Experimental Physics, Saarland University, 66041 Saarbr\"ucken, Germany}
% Experimental Physics, Saarland University, 66123 Saarbrücken, Germany
\author{Christian Wagner}
\affiliation{Experimental Physics, Saarland University, 66041 Saarbr\"ucken, Germany}
\affiliation{Physics and Materials Science Research Unit, University of Luxembourg, 1511 Luxembourg, Luxembourg}
% Physics and Materials Science Research Unit, University of Luxembourg, 162a avenue de la Faïencerie, L-1511 Luxembourg, Luxembourg
\author{Benjamin M. Friedrich}
\email{benjamin.m.friedrich@tu-dresden.de}
\affiliation{cfaed, TU Dresden, 01062 Dresden, Germany}

\date{\today}


\begin{abstract}
We present a theory of flagellar synchronization in the green alga \textit{Chlamydomonas}, using full treatment of flagellar hydrodynamics. 
We find that two recently proposed synchronization mechanisms, basal coupling and flagellar waveform compliance, stabilize anti-phase synchronization if operative in isolation. 
Their nonlinear superposition, however, stabilizes in-phase synchronization as observed in experiments. 
Our theory predicts different synchronization dynamics in fluids of increased viscosity or external flow, 
suggesting a non-invasive way to control synchronization by hydrodynamic coupling. %//600 charachters
\end{abstract}

\maketitle

\paragraph{Introduction.}
Pairs of coupled oscillators can synchronize with a fixed phase difference,
a phenomenon first observed by Huygens for a pair of pendulum clocks \cite{Pikovsky:synchronization}. 
Since then, synchronization has been described for many physical systems, 
including beating flagella \cite{Gray:1928}, pairs of heart muscle cells \cite{Nitsan:2016}, or light-driven micro-mills \cite{Kotar:2010}. 
In each of these different systems, 
the dynamics towards a synchronized state
is well approximated by the classic Adler equation 
for the phase difference $\delta$ between two weakly coupled oscillators \cite{Adler:1946,Stratonovich:1963}, 
which reads (for the simplest case of identical intrinsic frequencies $\omega_0=2\pi/T$ and absence of noise)
\begin{equation}
\label{eq_adler}
\dot{\delta} = -\frac{\lambda}{T} \sin\delta, \quad
\delta^\ast_\mathrm{IP}=0,\quad
\delta^\ast_\mathrm{AP}=\pi.
\end{equation}
The two steady states of Eq.~(\ref{eq_adler}), 
% $\delta^\ast=0$
$\delta^\ast_\mathrm{IP}$
and 
% $\delta^\ast=\pi$,
$\delta^\ast_\mathrm{AP}$, 
characterize in-phase synchronization (IP)
and anti-phase synchronization (AP), respectively, see Fig.~1(a,b).
The sign of the effective synchronization strength $\lambda$ selects
which state is stable.  Unless the oscillator coupling possesses
special symmetries, $\lambda$ is generically non-zero \cite{Elfring:2009,Friedrich:2016}. 
Its sign, however, depends on non-generic features of the system. 
For example, for a system of two beating metronomes on a moving tray -- a modern day analogue of Huygens' pendulum clocks -- 
both IP and AP synchronization were observed, depending on subtle features like
friction with the floor \cite{Pantaleone:2002}.
% last sentence could be shortened

\begin{figure}[h]
\begin{center}
\includegraphics[width=0.45\textwidth]{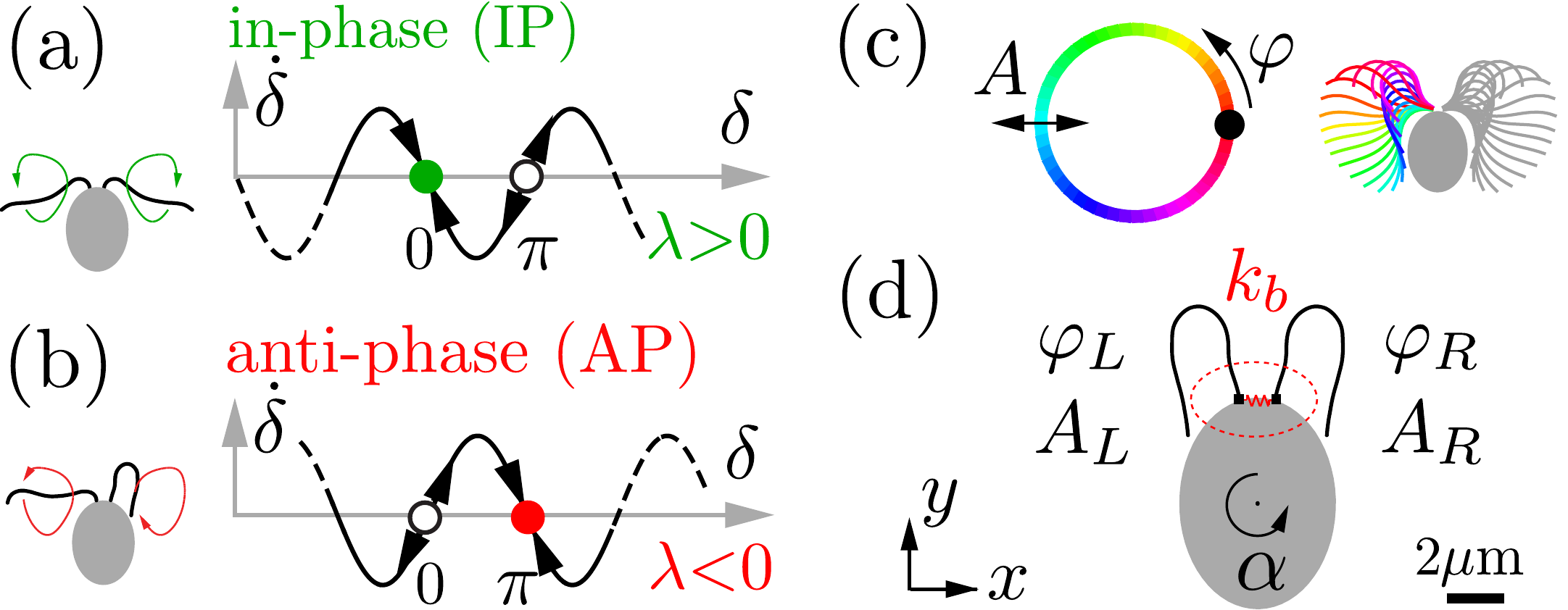}
\end{center}
\caption{In-phase and anti-phase synchronization. 
(a) The generic Adler equation, Eq.~(\ref{eq_adler}), 
predicts stable in-phase synchronization for a pair of coupled oscillators with phase difference $\delta=0$ for synchronization strength $\lambda>0$,
corresponding to ``breast-stroke swimming'' \textit{Chlamydomonas}.
(b) For $\lambda<0$, anti-phase synchronization is stable, corresponding to a ``free-style gait'' \cite{Wan:2016}.
(c) In our theory, we characterize each flagellum as a limit-cycle oscillator with phase $\varphi$ and amplitude $A$,
which uniquely determine flagellar shapes \cite{Klindt:2016}.
(d) We consider a generic elastic coupling between flagella bases with stiffness $k_b$, see Eq.~(\ref{eq_eb}).
}\label{fig:intro}
\end{figure}

At the microscopic scale of biological cells, cilia and flagella represent a prime example of a chemo-mechanical oscillator.
Molecular motors inside the flagellum drive regular bending waves of these slender cell appendages \cite{Alberts:cell},
rendering the flagellar beat a stable limit-cycle oscillator \cite{Ma:2014,Wan:2014,Klindt:2016}. 
% Single flagella can phase-lock to external oscillatory flows \cite{Quaranta:2015}.  
Pairs of flagella can synchronize their beat,
% for example in close-by swimming sperm cells \cite{Gray:1928,Woolley:2009}, or 
e.g.\ in the green alga \chlamy\ that swims with $n=2$ flagella like a breast-stroke swimmer
\cite{Ruffer:1998a,Goldstein:2009,Geyer:2013}. 
IP synchrony of its two flagella is a prerequisite for swimming straight and fast.
The basal bodies of the two flagella are connected by a so-called distal striated fiber \cite{Ringo:1967}.
More complex flagellar gaits were observed in species with $n=2^k$ flagella,
with matching patterns of basal coupling \cite{Wan:2016}. 
On epithelial surfaces, $n\gg 10^2$ flagella phase-lock their beat,
thus forming metachronal waves \cite{Sanderson:1981}, 
% resulting in metachronal waves of coordinated flagellar beating \cite{Sanderson:1981}, 
which facilitates efficient fluid transport \cite{Cartwright:2004,Osterman:2011,Elgeti:2013}.  
Flagellar synchronization has been studied intensively in the model organism \textit{Chlamydomonas}, 
reporting both IP and AP beating.
While wild-type \textit{Chlamydomonas} cells usually
display IP synchrony, stochastic switching between regimes of stable
IP and AP beating has been observed in a flagellar mutant (ptx1) % non-phototactic, flagellar dominance mutant (cis and trans respond alike)	
\cite{Leptos:2013}.  Another mutant (vfl) with impaired basal coupling displayed lack of coordinated flagellar beating altogether \cite{Quaranta:2015,Wan:2016}.

A long-standing hypothesis states that flagellar synchronization arises from a hydrodynamic coupling between flagella \cite{Taylor:1951},
as demonstrated for pairs of flagellated cells held at a distance \cite{Brumley:2014}. 
A popular minimal model of this phenomenon abstracts from the specific shape of flagellar bending waves
and represents each flagellum by a sphere moving along a circular orbit \cite{Vilfan:2006,Niedermayer:2008,Uchida:2011,Friedrich:2012c,Bennett:2013,Polotzek:2013,Izumida:2016}. 
The motion of the left and the right sphere with respective phase angles $\varphi_L$ and $\varphi_R$ is described by a balance of forces
between active driving forces $Q_j$ and hydrodynamic friction forces
\begin{equation}
\label{eq:two_spheres}
Q_j = \Gamma_{jL}\dot{\varphi}_L + \Gamma_{jR}\dot{\varphi}_R,\quad
j\in\{L,R\}.
\end{equation} 
Specifically, $\Gamma_{LL}\dot{\varphi}_L$ is the hydrodynamic friction force acting on the left sphere due to its own motion,
and $\Gamma_{LR}\dot{\varphi}_R$ represents direct hydrodynamic interactions exerted by the right sphere on the left one.
% This minimal model highlights the role of symmetry-breaking for synchronization  \cite{Elfring:2009,Elgeti:2015,Friedrich:2016}. 
The minimal model possesses parity-time symmetry (PT),
characterized by $\Gamma_{LR}(\varphi_L,\varphi_R)=\Gamma_{RL}(-\varphi_R,-\varphi_L)$, 
i.e.\ a spatial parity transformation ($\varphi_L\leftrightarrow -\varphi_R$) gives rise to an equivalent dynamics, but with time-arrow reversed 
\cite{Elfring:2009,Elgeti:2015,Friedrich:2016}.
Thus, there can be neither stable nor unstable states, unless PT-symmetry is broken.

A number of different PT-symmetry breaking effects have been proposed in the past, 
including interaction with boundary walls \cite{Vilfan:2006}, 
phase-dependent driving forces $Q_j(\varphi)$ \cite{Uchida:2011}, 
% finite Reynolds number effects \cite{Theers:2013}, 
and amplitude compliance with a variable radius $A(t)$ of each circular orbit, constrained by an elastic spring \cite{Niedermayer:2008,Reichert:2005}. 
In addition to direct hydrodynamic interactions between the two flagella,
synchronization independent of hydrodynamic interactions can occur by a coupling between flagellar beating and the resultant motion of the cell \cite{Friedrich:2012c}.
Importantly, two recent experimental studies suggest that in \textit{Chlamydomonas},
an elastic basal coupling connecting the proximal ends of both flagella  
could play a key role for flagellar synchronization \cite{Quaranta:2015,Wan:2016}. 
While each of these proposed mechanisms could in principle account for synchronization, it is not known, which symmetry breaking mechanism dominates in the real biological system.
\textit{A priori}, we do not even know if a specific mechanism will stabilize the IP or AP synchronized state.

Here, we theoretically study flagellar synchronization in the model organism \chlamy\ to predict conditions for IP and AP synchrony,
and present a first experiment to test our theory.
We build on a previously developed description of the beating flagellum as a limit-cycle oscillator \cite{Klindt:2016}. 
% , which was calibrated and tested against experimental data \cite{Klindt:2016}. 
There, we retain the picture of a point moving along a circular orbit, yet each position of
this point represents a genuine flagellar shape, see Fig.~\ref{fig:intro}(c). % \cite{Werner:2014,Ma:2014}. 
% Thus, any motion of this point with polar coordinates $(\varphi,A)$ corresponds to a shape
% change of the flagellum, for which we numerically compute hydrodynamic
% friction forces acting on the cell and its two flagella. 
Our theory uses detailed hydrodynamic calculations based on experimental beat patterns, to elucidate two PT-symmetry breaking effects: 
flagellar waveform compliance, and
% by introducing a flagellar amplitude $A$ that describes flagellar shape changes in a two-dimensional space of principal flagellar shape modes 
% determined from experimental data \cite{Klindt:2016}.  
basal coupling between both flagella.  
We find that both PT-symmetry breaking mechanisms have a strong impact on synchronization, 
but only their combination yields IP synchrony with a synchronization strength sufficient to overcome noise \cite{Goldstein:2009,Ma:2014}.
 
\paragraph{Effective theory of flagellar swimming and synchronization.}
Recently, we introduced an effective theory of flagellar swimming \cite{Klindt:2016},
there formulated for the case of synchronized beating only.
The main idea behind this theory is that regular flagellar bending waves
represent a limit-cycle oscillator, which can be generically parametrized 
by a $2\pi$-periodic phase variable $\varphi$ obeying $\dot{\varphi}=\omega_0$ in the absence of an external perturbation, 
as well as a normalized amplitude $A$, which will always relax to a steady-state value $A_0$.
Any deviation from the reference condition, 
e.g.\ for asynchronous beating, external flow, or altered viscosity of the surrounding fluid, 
changes $\dot{\varphi}$ or $A$. 
We can thus describe the motion of a \chlamy\ cell in a plane by a state vector $\q$ comprising seven degrees of freedom,
see Fig.~\ref{fig:intro}(d)
\begin{equation}
\q= ( \varphi_L, A_L, \varphi_R, A_R, \alpha, x, y )^T.
\end{equation}
Here, $\varphi_j$, $A_j$ with $j\in\{L,R\}$ denote phase and amplitude
of the left and right flagellum, respectively, while $\alpha$, $x$, $y$,
denote orientation angle and center position of the cell body.

Each change of a degree of freedom will set the surrounding fluid in motion and induce hydrodynamic dissipation, 
in addition to friction inside the flagella.
The total dissipation rate $\R$ can be expressed in terms of generalized velocities $\dot{q}_j$ and conjugated generalized friction forces $P_j$ 
\begin{equation}
\R = P_{\varphi_L} \dot{\varphi}_L + P_{A_L} \dot{A}_L + P_{\varphi_R} \dot{\varphi}_R + P_{A_R} \dot{A}_R + P_\alpha \dot{\alpha} + P_x \dot{x} + P_y \dot{y}.
\end{equation}
The definition of the generalized friction forces $P_j$ follows
the framework of Lagrangian mechanics for dissipative systems, using
$\R$ as Rayleigh dissipation function \cite{Goldstein:mechanics,Polotzek:2013}. 
In the limit of zero Reynolds number, applicable to cellular
self-propulsion where inertia is negligible \cite{Lauga:2009}, 
hydrodynamic friction forces are linear in the generalized velocities $q_j$,
$P^{(h)}_i=\Gamma_{ij}^{(h)}\dot{q}_j$ (Einstein summation convention).
% due to the linearity of the Stokes equation. 
The total friction forces $\P=\P^{(h)}+\P^{(i)}$ additionally comprise intraflagellar friction forces $\P^{(i)}$.
For simplicity, we assume $P^{(i)}_i=\Gamma_{ij}^{(i)}\dot{q}_j$ with coefficients proportional to the respective hydrodynamic friction coefficients,
i.e.\
$\Gamma_{ij}^{(i)} = (1-\eta)/\eta\, \Gamma_{ij}^{(h)}$ 
for either $i,j\in\{\varphi_L,A_L\}$ or $i,j\in\{\varphi_R,A_R\}$ and $\Gamma_{ij}^{(i)}=0$ else, 
where $\eta$ denotes an energy efficiency of the flagellar beat.

We coarse-grain the activity of molecular motors inside each flagellum 
by active flagellar driving forces $Q_{\varphi_j}(\varphi_j)$ and 
amplitude restoring forces $Q_{A_j}(\varphi_j)$, $j\in\{L,R\}$.
Thus, at each instance in time, we have 7 force balance equations
\begin{equation}
\label{eq_PQ}
Q_j = P_j, \quad j\in\{\varphi_L,A_L,\varphi_R,A_R,\alpha,x,y\}.
\end{equation}
Here, the generalized forces $Q_x$, $Q_y$, and $Q_\alpha$ represent constraining forces that ensure 
constraints of motion imposed on the cell.
For a freely-swimming cell, force and torque balance imply $Q_x=Q_y=0$, $Q_\alpha=0$. 
For a fully clamped cell, one would impose $\dot{x}=0$, $\dot{y}=0$, $\dot{\alpha}=0$, 
and determine the constraining forces $Q_x$, $Q_y$, $Q_\alpha$ such that the constraints are satisfied.
% Below, we also consider the case of an elastically clamped cell with  $\dot{x}=0$ and $\dot{y}=0$ that is subject to an elastic restoring torque $Q_\alpha = - k_\alpha \alpha$ with some torsional stiffness $k_\alpha$.
With this calibration, Eq.~(\ref{eq_PQ}) fully specify equations of motions of flagellar swimming and synchronization.

Hydrodynamic computations allow us to determine all hydrodynamic
friction coefficients $\Gamma^{(h)}_{ij}$ for a given flagellar beat
pattern.  Here, we use a fast boundary element method as described in
\cite{Klindt:2015} and a flagellar beat pattern recorded for the
reference condition of a clamped cell with IP-synchronized beat
\cite{Klindt:2016}.  There, the efficiency parameter has been
estimated as $\eta=0.21\pm0.06$ \cite{Klindt:2016}.
Additionally, the flagellar driving forces were uniquely calibrated from the requirement
$\dot{\varphi}_L=\dot{\varphi}_R=\omega_0$ and $A_L=A_R=A_0$ for the reference case.
The amplitude restoring forces
$Q_{A_j}$ determine how fast amplitude perturbations $A-A_0$ decay.
Here, we assume exponential relaxation with a single relaxation time-scale $\tau_A$ for the reference condition, 
which uniquely determines $Q_{A_j}$ \cite{Klindt:2016}.
For $\omega_0\tau_A\ll 1$, perturbations cannot change the amplitude,
while  for $\omega_0\tau_A\gg 1$ the limit cycle may become unstable.
An analysis of amplitude fluctuations of the flagellar beat provided an estimate
$\tau_A\approx 6\,\mathrm{ms}$ \cite{Ma:2014}. 
We now use this theoretical
description to predict dynamics after a perturbation of perfect synchrony for
different PT-symmetry breaking scenarios.

\paragraph{Flagellar waveform compliance.}
Elastic degrees of freedom such as a flagellar waveform compliance 
can break PT symmetry in minimal models of hydrodynamically coupled oscillators, 
and thus allow for synchronization \cite{Niedermayer:2008}.
We tested this general proposition for the specific case of flagellar synchronization in \chlamy, 
using our theoretical description with amplitude degrees of freedom $A_L$ and $A_R$. 
We quantify the stability of the IP-synchronized state in terms of an effective synchronization strength $\lambda$, 
generalizing the parameter $\lambda$ in Eq.~(\ref{eq_adler}), 
such that $-\lambda/T$ equals the cycle-average Ljapunov exponent for the phase difference $\delta=\varphi_L-\varphi_R$.
The sign of $\lambda$ indicates whether IP synchrony is stable ($\lambda>0$) or not ($\lambda<0$).

\begin{figure}
\begin{center}
\includegraphics[width=0.5\textwidth]{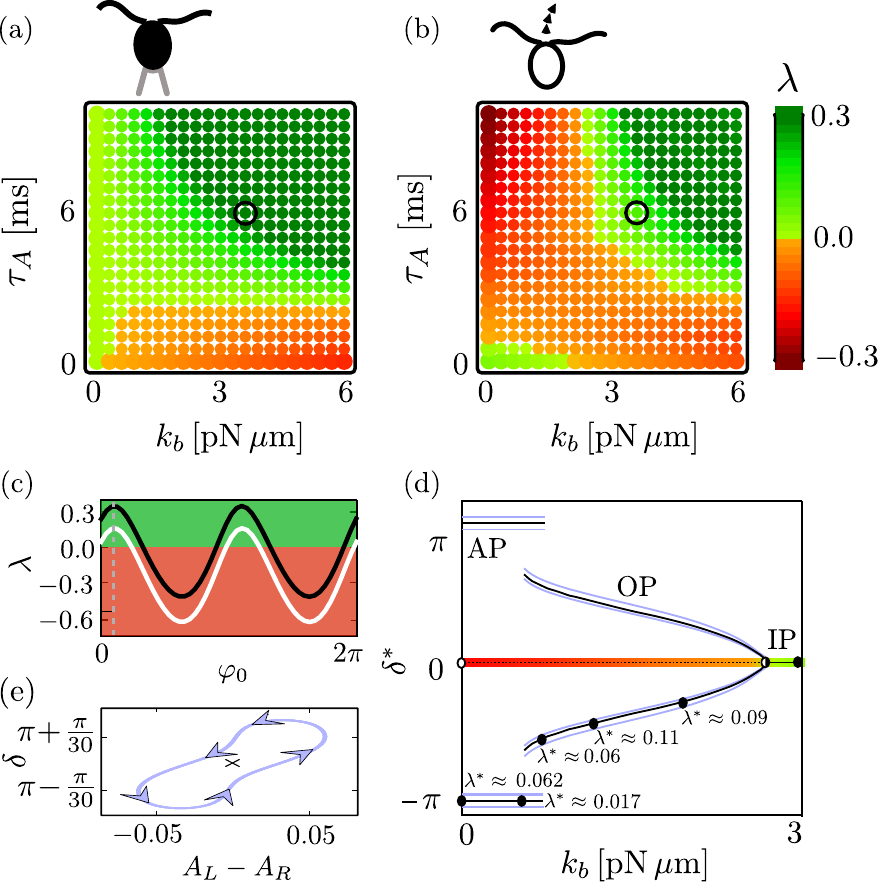}
\end{center}
\caption[]{ 
Waveform compliance and basal coupling jointly determine synchronization dynamics in \chlamy.  
(a) Computed synchronization strength $\lambda$ (color code) 
as function of amplitude relaxation time $\tau_A$ and basal coupling stiffness $k_b$ for a clamped cell.
(b) Same for a free-swimming cell. 
(c) $\lambda$ becomes maximal at $\varphi_0\approx \pi/10$ for both clamped (black) and free-swimming cells (white). 
(d) Spontaneous symmetry breaking of synchronization: 
Cycle-averaged phase difference $\delta^\ast$ at steady state as function of basal stiffness $k_B$ computed for a free-swimming cell (black). 
For selected steady states, we computed synchronization strengths $\lambda^\ast$ characterizing Ljapunov exponents $-\lambda^\ast/T$ of convergence towards $\delta^\ast$. 
Blue lines indicate maximum and minimum values of small-amplitude oscillations around $\delta^\ast$ at steady state,
see also panel (e) [$k_b=0$]. 
Parameters:
$T=20\,\mathrm{ms}$,
$\tau_A = 6\,\mathrm{ms}$,
$k_B = 3.6\,\mathrm{pN\,\mu m}$, 
$\eta=0.2$, 
$\mu=0.85\, \mathrm{mPa\,s}$,
$\varphi_0 = \pi/10$,
unless stated otherwise.
}
\label{fig:basal_coupling}
\end{figure}

We computed $\lambda$ for both the case of free-swimming and of clamped cells
for two waveform data sets,
see Fig.~\ref{fig:basal_coupling} and Fig.~S4 in SM for $k_b=0$ (no basal coupling).
Whether a cell can swim freely, or is restrained from moving, can make a substantial difference for flagellar synchronization \cite{Geyer:2013}.
In the absence of flagellar waveform compliance ($\tau_A=0$) and basal coupling ($k_b=0$),
we find $\lambda\approx 0.06$ for a free-swimming cell, and $\lambda\approx 0$ for a clamped cell,
similar to a previous study \cite{Geyer:2013}
\endnote{
%Ref.~\cite{Geyer:2013} used 
There, $\eta=1$, implying $\lambda\approx 0.3$, see Fig.~S5 in SM.
}.
Amplitude compliance ($\tau_A>0$) changes the synchronization strength, 
yet, surprisingly, destabilizes IP synchrony for free-swimming cells. 
Next, we study how an elastic basal coupling affects flagellar synchronization.

\paragraph{Basal body coupling.}
In \chlamy, the proximal ends of both flagella are connected by a distal striated fiber, 
comprising an elastic basal coupling \cite{Ringo:1967}.
Previous experimental studies indicate the importance of this basal link for flagellar synchronization \cite{Quaranta:2015,Wan:2016}. 
In the following, we account for a finite elastic stiffness of this basal link, 
for which we assume a Hookian elastic energy 
\begin{equation}
\label{eq_eb}
U_b = \frac{k_b}{2 b_1^2} [ b(\varphi_L, A_L, \varphi_R, A_R)-b_0 ]^2.
\end{equation}
Here, $b(\varphi_L, A_L, \varphi_R, A_R)$ represent the elongation of
the basal link, which is a periodic function of the two
flagellar phases.  In the absence of detailed knowledge of the elastic
properties of the basal apparatus, we make the generic Ansatz
$b(\varphi_L, A_L, \varphi_R, A_R) = b_0+b_1 A_L\sin(\varphi_L-\varphi_0)+b_1 A_R\sin(\varphi_R-\varphi_0)$
with some phase shift $\varphi_0$. 
The elastic energy of the basal link results in an additional term in the
active flagellar driving force
\begin{equation}
\label{eq_Qb}
  Q_{\varphi_L} \rightarrow Q_{\varphi_L}
  - \frac{\partial U_b}{\partial \varphi_L}_{|\varphi_L, A_L,
    \varphi_R, A_R}
  + \frac{\partial U_b}{\partial \varphi_L}_{|\varphi_L, A_L,
    \varphi_L, A_L},
\end{equation}
and similarly for $A_L$, $\varphi_R$, $A_R$.  Here, the last term
merely reflects the fact that the elastic basal coupling must be
incorporated in the calibration of the flagellar driving forces to
yield $\dot{\varphi}_L=\dot{\varphi}_R=\omega_0$ in the reference case
of IP-synchronized beating. 
Note that the unknown phase shift $\varphi_0$ affects synchronization, see Fig.~2(c).

Fig.~\ref{fig:basal_coupling}(a,b) shows numerical results for the
synchronization strength $\lambda$ as a function of basal stiffness $k_b$ 
% for a particular value of $\varphi_0$ (such that $\lambda$ is maximal)
for both clamped and free-swimming cells.
Remarkably, basal coupling destabilizes IP synchrony in the absence of amplitude compliance, 
but stabilizes it for realistic values of the amplitude relaxation time $\tau_A$ and suitable choice of $\varphi_0$.
Thus, the combined effect of two PT-symmetry breaking mechanisms is opposite to the sum of their individual effects.
A basal stiffness of
$k_b= 3.6\,\mathrm{pN\,\mu m}$
reproduces a previously measured value of
$\lambda \approx 0.3$ for clamped cells~\cite{Goldstein:2009}.
With the length $300\,\mathrm{nm}$ and cross-sectional area $2\cdot 10^4\,\mathrm{nm}^2$ 
of the distal striated fiber \cite{Ringo:1967},
and assuming $b_1=50\,\mathrm{nm}$,
our estimate for $k_b$ corresponds to a Young's modulus of approximately $20\,\mathrm{kPa}$,
well in the range of biological materials.

\paragraph{Out-of-phase synchronization.} 
Flagellar synchronization by basal coupling exhibits dynamics that is more complex then the Adler equation.
% Fig.~\ref{fig:basal_coupling}(d) displays the cycle-averaged phase difference $\delta^\ast$ between both flagella at steady state as a function of basal stiffness $k_b$. 
While we find stable AP and IP synchronization for sufficiently weak and strong basal coupling, respectively,
consistent with Eq.~(\ref{eq_adler}),  
we find a regime of out-of-phase (OP) synchronization with $0<\delta^\ast<\pi$ for intermediate coupling strengths,
emerging from the IP-synchronized state by a pitchfork bifurcation, % double-check nomenclature
see Fig.~\ref{fig:basal_coupling}(d).
This OP synchronization represents an instance of spontaneous symmetry-breaking with two stable solutions $\pm\delta^\ast$. 

\begin{figure}
\begin{center}
\includegraphics[width=0.5\textwidth]{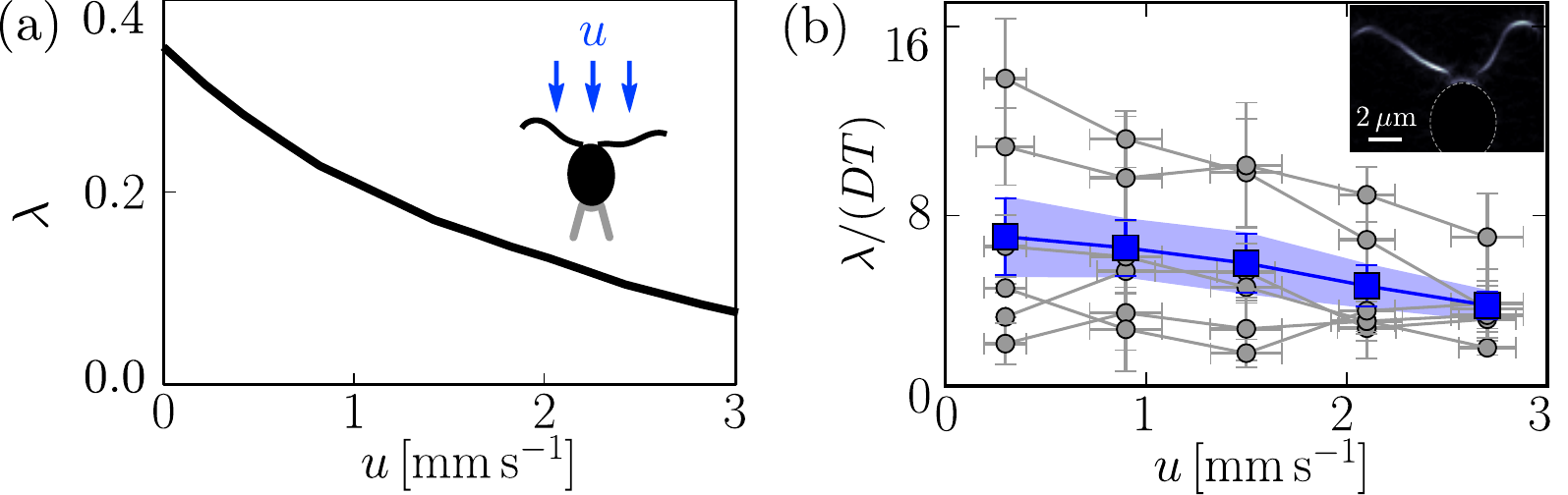}
\end{center}
\caption[]{
Control of flagellar synchronization by external flow.
(a) Theory: computed synchronization strength $\lambda$ as function of external fluid flow with velocity $u$ parallel to the long axis of the cell.  
(b) Experiment: measured synchronization strength $\lambda/(DT)$, normalized by effective noise strength $D$
(blue: mean$\pm$s.e.m., $n=6$ cells; gray: mean$\pm$s.e.\ for individual cells).
Parameters: see Fig.~\ref{fig:basal_coupling}.
}
\label{fig:exp}
\end{figure}

\paragraph{Suggestions for experiments.}
Our theory suggests a non-invasive way to control flagellar synchronization. 
We predict that for external flow parallel to the long axis of a \chlamy\ cell, 
the synchronization strength is reduced, see Fig.~\ref{fig:exp}(a).
Increasing the viscosity of the surrounding fluid gives similar results, see Fig.~S6 in the Supporting Material (SM).
Conceptually, 
imposing an external flow is equivalent to changing the phase-dependence of the flagellar driving forces, 
while increasing the viscosity reduces the magnitude of elastic coupling relative to viscous coupling.

We performed experiments, exposing \textit{Chlamydomonas} cells held in micropipettes to external flow.
We determined a flow-dependent synchronization strength $\lambda/(DT)$,
normalized by an effective noise strength $D$ of flagellar beating, 
see Fig.~\ref{fig:exp}(b) and SM for details.  
Independent measurements reported $DT\approx 0.1-0.2$ \cite{Goldstein:2011,Ma:2014}. 
This suggests a quantitative match of theory and experiment.

\paragraph{Minimal model of synchronization by basal coupling.}
To gain insight into basic mechanisms of IP and AP synchronization, 
we revisit a popular minimal model of hydrodynamic synchronization \cite{Vilfan:2006,Niedermayer:2008,Uchida:2011}.
Two spheres of equal radius $r$ move inside a viscous fluid of viscosity $\mu$ along circular orbits $\r_j$ of respective radii $A_j$,
with centers separated by a distance $d$,  
$\r_j=A_j\n_j + \sigma_j d \e_x/2$, for $j\in\{L,R\}$.
Here, 
$\n_j=\cos\varphi_j\,\e_x+\sigma_j\sin\varphi_j\,\e_y$ denotes the radial vector and $\sigma_L=-1$, $\sigma_R=1$.
Each sphere is driven by a constant tangential driving force
$Q_j = q_0$ with $q_0=A_0^2 \gamma \omega_0$, friction coefficient $\gamma=6\pi\mu r$ and reference amplitude $A_0$.
Hydrodynamic interactions couple the motion of both spheres.
In the limit $r{\ll}d$ with $A_0/r$ of order unity,
$\Gamma_{LR}\dot{\varphi}_R=-A_L\gamma^2\,\t_L\cdot\mathcal{G}(\r_L-\r_R)\cdot\dot{\r}_R$, and vice versa.
Here, $\t_j=\partial \n_j/\partial\varphi_j$ is the tangent vector and 
$\mathcal{G}(\r)=(8\pi\mu)^{-1} [|\r|^{-1}+\r\otimes\r/|\r|^3] $ denotes the Oseen tensor.
For constant amplitude, $A_i=A_0$, the system possesses PT-symmetry and no net synchronization occurs \cite{Vilfan:2006,Elfring:2009,Friedrich:2016}.
Introducing amplitude compliance, 
$\gamma\dot{A}_L = -k_A(A_L-A_0) - \gamma^2 \n_L\cdot\mathcal{G}(\r_R-\r_L)\cdot\dot{\r}_R$
with amplitude stiffness $k_A$
for the left sphere and similarly for the right sphere, 
breaks PT-symmetry and results in
$\lambda_a=-3\pi\,\tau_A\omega_0 r/(4d)+\mathcal{O}(r/d)^2$, 
where $\tau_A=\gamma/k_A$ denotes an amplitude relaxation time, 
see SM for details. 
Note that we consider counter-rotating spheres,   
mimicking a clamped \textit{Chlamydomonas} cell \cite{Friedrich:2012c,Bennett:2013}, 
while the originally studied case of co-rotating spheres yields
$\lambda_a=9\pi\,\tau_A\omega_0 r/(2d)+\mathcal{O}(r/d)^2$ \cite{Niedermayer:2008}.
Analogous to Eq.~(\ref{eq_Qb}), 
we can introduce `basal coupling' in this two-sphere model 
[with $U_b=k_b |\r_R-\r_L|^2/(2 A_0^2)$]
as a second PT-symmetry breaking mechanism. 
This yields a synchronization strength
$\lambda_b=-\pi\,k_b/q_0+\mathcal{O}(r/d)$ in the absence of amplitude compliance with $\tau_A=0$.
Thus, both mechanism imply $\lambda<0$ for $r,A_0\ll d$ if operative in isolation.
Their nonlinear superposition, however, results in a positive cross-coupling term
\begin{equation}
\lambda=\lambda_a+\lambda_b+\frac{\pi}{2}\,
(k_b/q_0)^2\,
% \frac{k_b^2}{q_0^2}\,
\tau_a\omega_0+\mathcal{O}(r/d) .
\end{equation}
As a consequence, IP synchrony is stable for suitable $k_b>0$ and $\tau_A>0$.
% Higher-order terms depend on $\varphi_0$.

\paragraph{Discussion.}
Here, we presented a theory of flagellar swimming and synchronization for the model organism \chlamy, 
to dissect the role of two proposed synchronization mechanism, flagellar waveform compliance \cite{Niedermayer:2008} and elastic basal coupling \cite{Quaranta:2015,Wan:2016}.
We find that each mechanism separately stabilized anti-phase synchronization in free-swimming cells, 
but their combination results in in-phase synchronization, as observed in experiments \cite{Ruffer:1998a,Goldstein:2009}.

Our theory makes specific predictions that can be tested in experiments.
This includes altered synchronization dynamics in the presence of external flow or fluids of increased viscosity. 
Further, experimental disruption of the distal striated fiber that link the basal bodies of the two flagella, 
e.g.\ by laser ablation, could validate the role of basal coupling for synchronization proposed here.
Interestingly, a change in length of the distal striated fiber, e.g.\ induced by intracellular calcium signaling \cite{Salisbury:1978}, 
could allow the cell to switch between IP and AP synchronization (see Fig.~S7 in SM),
causing a `run-and-tumble' motion as observed previously \cite{Polin:2009}. 
In conclusion, we have shown that synchronization strengths measured in experiments \cite{Goldstein:2009}
cannot be explained in our theory without basal coupling, 
yet are reproduced for plausible parameter choices assuming such coupling.

\paragraph{Acknowledgment.}
G.S.K. and B.M.F. acknowledge support from
the German Science Foundation ``Microswimmers" Priority Program 1726 (Grant No. FR 3429/1-1). 

\bibliography{E:/ben/bibliography/library}

\end{document}